# Computational and Experimental Exploration of Protein Fitness Landscapes: Navigating Smooth and Rugged Terrains

Mahakaran Sandhu*[1,2], John Chen*[1,2], Dana Matthews*[1,2], Matthew A Spence[1,2], Sacha B Pulsford[1,2], Barnabas Gall[1,2], James Nichols[3], Nobuhiko Tokuriki[4], Colin J Jackson[1,2,3,5]

1 Research School of Chemistry, Australian National University, Canberra, ACT 2601, Australia

2 Centre of Excellence for Innovations in Peptide & Protein Science, Research School of Chemistry, Australian National University, Canberra, ACT 2601, Australia

3 Research School of Biology, Australian National University, Canberra, ACT 2601, Australia

4 Michael Smith Laboratories, University of British Columbia, Vancouver, Canada

5 ARC Centre of Excellence in Synthetic Biology, Research School of Biology, Australian National University, Canberra, ACT 2601, Australia

To Whom Correspondence Should be Addressed: colin.jackson@anu.edu.au

## Abstract

Proteins evolve through complex sequence spaces, with fitness landscapes serving as a conceptual framework that links sequence to function. Fitness landscapes can be smooth, where multiple similarly accessible evolutionary paths are available, or rugged, where the presence of multiple local fitness optima complicate evolution and prediction. Indeed, many proteins, especially those with complex functions or under multiple selection pressures, exist on rugged fitness landscapes. Here we discuss the theoretical framework that underpins our understanding of fitness landscapes, alongside recent work that has advanced our understanding - particularly the biophysical basis for smoothness *versus* ruggedness. Finally, we address the rapid advances that have been made in computational and experimental exploration and exploitation of fitness landscapes, and how these can identify efficient routes to protein optimization.

## Introduction

Proteins evolve *via* mutation and selection within vast multidimensional sequence spaces. Selection acts to purge or retain sequences (genotype) based on their fitness (phenotype),[1] a term that can encompass properties such as catalytic efficiency,[2] binding affinity,[3] stability,[4] or any functional trait relevant to a given biological context. Protein fitness is the most direct trait for understanding molecular evolution and for applications to protein engineering. However, in natural evolution, protein fitness can only be selected for as a function of its contribution to organism fitness (survival). Both levels of fitness are linked and jointly affect the evolutionary dynamics. The concept of the fitness landscape is used to map the relationship between sequence and function, where each protein sequence is a “coordinate” with a corresponding fitness value.

This perspective is mostly focused on protein fitness. The concept of the protein fitness landscape is used to map the relationship between sequence and function, where each protein sequence is a “coordinate” with a corresponding fitness value. This offers a powerful framework for understanding evolutionary dynamics in proteins,[5,6] as traversal of the landscape via mutational steps can be used to model and dissect factors that drive the evolutionary process. In this framework proteins can evolve by taking mutational steps across the landscape toward fitness “peaks” or “optima”, depending on the selection pressure. The topology of the landscape determines the accessible paths that any given evolutionary trajectory (the mutational path from starting to end sequence) can take, usually with the constraint that fitness must increase (or at least remain with a “neutral” range) at every step. Key topological features that shape evolution include fitness “valleys”, regions of low fitness that cannot be traversed and must be circumvented in some way, and the existence of multiple optima, which may cause an exploring sequence to be trapped on a local optimum. Landscapes that feature such complexities are termed “rugged”, while those that are relatively simple to traverse are considered “smooth”. Smoothness is a consequence of simple additive mutational effects,[7,8] whereas ruggedness can be caused, at least in part, by mutational epistasis:[5,7,9] the phenomenon whereby the effect of a mutation is dependent on the sequence context,[10] i.e. a mutation could be beneficial or deleterious in different sequence backgrounds.

Fitness landscapes—whether smooth or rugged—dictate how proteins can evolve over time, either through natural selection or laboratory-driven protein engineering and design efforts.[11] The ruggedness of fitness landscapes is shaped by the underlying biophysical and structural properties of proteins, and has important consequences for protein engineering, evolution, and sequence-fitness prediction using machine learning (ML). Understanding these landscapes is crucial for elucidating evolutionary dynamics and improving strategies for rational and ML-guided protein design.[11,12]

### Glossary

**Evolutionary contingency:** The dependence of the evolutionary trajectory on its starting sequence.
**Compensatory mutations:** Mutations that compensate for deleterious mutations, e.g. a stabilizing mutation could “rescue” a protein that cannot fold efficiently.
**Fitness graph:** A graph in which the vertices are associated with sequence(s), and edges connect sequences depending on a distance metric (e.g., mutational or “Hamming” distance).
**Graph Fourier Transform (GFT):** The representation of a signal on a graph (e.g., a fitness landscape) as a linear combination of eigenvectors of the graph Laplacian. Analogous to the continuous Fourier Transform, more rugged landscapes have a higher contribution from higher frequency eigenvectors and vice versa.
**Graph Laplacian:** A matrix representation of a graph that captures its structure and properties.
**Hamming distance:** The number of positions at which two equal-length strings (protein sequences) are different. A measure of mutational distance.
**Local maxima:** A sequence whose immediate neighbors all have lower fitness values.
**Global maxima:** The sequence with the highest fitness in the entire landscape.
**Neutral drift:** Exploration of mutations that are neither deleterious nor advantageous with respect to fitness but affect what subsequent mutations are accessible to the protein.
**Adaptive walk:** The series of mutations that a protein undergoes as it climbs toward higher fitness.

## Smooth fitness landscapes

The "smoothness" of a protein fitness landscape reflects the accessibility of the fitness optimum to any given sequence in sequence space (**Figure 1A**).[13] Mechanistically, the effects of mutations on a smooth landscape are additive and largely independent of each other (i.e., little to no epistasis), meaning beneficial mutations are beneficial regardless of the sequence background they are introduced into. Such a landscape is said to be smooth because the fitness maximum (i.e. the most fit sequence) can be accessed from many coordinates in the landscape via the gradual accumulation of beneficial mutations along a linear fitness gradient. By extension, in a smooth landscape the effects of multiple mutations are predictable: the fitness effects of combinatorial mutations are the sum (i.e., linear combination) of each individual mutational effect. However, the fitness landscapes of most real proteins are complicated by the effects of selection pressures on multiple protein traits (e.g., solubility, activity, folding) that simultaneously contribute to overall fitness. This can manifest as epistasis, where the fitness of a combinatorial mutant cannot be fully described as a linear combination of constituent mutations. This is especially true when we consider high-order interactions over all residues in the sequence. Despite the almost ubiquitous presence of some level of epistasis and ruggedness, real protein landscapes with a single fitness peak,[14] and landscapes where multiple mutational paths to the global fitness maximum are accessible,[13,15,16] have been observed and studied.

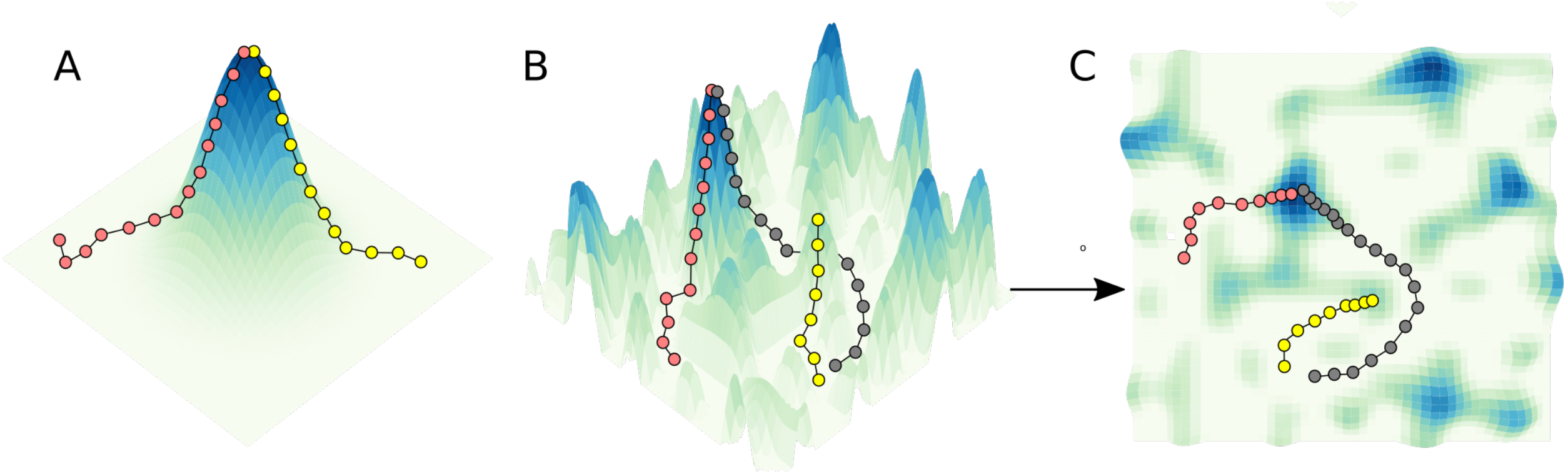

**Figure 1. The characteristics of evolutionary trajectories on smooth versus rugged landscapes are distinct.** (A) On a smooth landscape with a single peak, sequences can begin anywhere on the landscape and evolve to reach the same peak. (B) On a rugged landscape, the ability to reach a given peak is contingent on the position of the sequence on the fitness landscape due to the presence of multiple local maxima separated by fitness valleys. For example, the red evolutionary trajectory reaches the global maxima, while the yellow trajectory ascends a local fitness maximum. The gray trajectory, although beginning in a similar position as the yellow trajectory, undergoes neutral drift through low-fitness regions, accumulating neutral mutations that ultimately position it at the bottom of the global maximum, permitting it to ascend. (C) A "bird's-eye" perspective of the same landscape as B.

## Rugged fitness landscapes

In contrast to smooth landscapes, rugged protein fitness landscapes present a much more complex evolutionary terrain (**Figure 1B, C**).[17] These landscapes are characterized by multiple fitness peaks and valleys, where proteins may become trapped at local optima,[18] making further fitness improvements difficult. As a result, evolutionary contingency emerges as a major force shaping evolutionary trajectories, meaning that fitness maxima can be ascended only from certain coordinates on the landscape. The starting position of an evolutionary trajectory (i.e. the starting sequence) thereby determines which fitness maxima are accessible. The presence of multiple local optima means that even though a protein may be functional, its current sequence may prevent it from achieving higher fitness without first acquiring mutations that initially reduce its fitness.[19,20] Furthermore, it is possible for the same starting point to reach different optima based on the initial mutational steps, i.e. the first step, or mutation, determines which one local maxima becomes accessible.[21,22] These phenomena make evolutionary pathways through rugged landscapes complex and difficult, although not impossible, to predict.[23]

**The effects of epistasis on fitness landscapes**

A key driver of ruggedness is epistasis i.e., the dependence of the effects of a mutation on the sequence into which it is introduced. In rugged landscapes, interactions between mutations are complex and non-additive, resulting in fitness outcomes that are unpredictable. Epistasis can be divided into two categories, "nonspecific", and "specific".[10,24] In nonspecific epistasis, the mutational effects at the molecular level could be additive (non-epistatic) but appear to be epistatic due to a non-linear relationship between molecular level (stability) and higher-level effects (activity or fitness).[25–27] For example, a mutation could appear epistatic if an increase in activity was offset by a decrease in soluble protein expression due to reduced stability. Indeed, non-specific epistasis can emerge as a function of pleiotropy, where the same mutation affects multiple phenotypes (e.g., both stability and activity).[28] Due to this, epistasis is more likely to arise in proteins that have multiple interacting functional features, such as allosteric regulation, multiple interacting functional domains, as well as protein-protein and protein-nucleic-acid interactions.[24,29,30]

If the non-linear mapping is known, nonspecific epistasis can still produce somewhat predictable landscapes. In contrast, specific epistasis is the result of mutational interactions leading to non-additive effects, causing complex interdependencies between residues that lead to fitness peaks separated by fitness valleys, where a given beneficial mutation may be inaccessible because it requires traversal of a region of lower fitness.[31] Within a single protein, specific epistasis is more prevalent between residues that directly interact with each other in physical space and are involved in collectively determining protein function residues (e.g., in or near the active site of an enzyme).[32]

Owing to the expansive nature of sequence space, it is challenging to comprehensively explore epistasis in a full combinatorial protein pace ($20^N$ sequences for protein of length N). In practice, we are interested in the landscape features of local regions of the full space, typically around homologous proteins and ancestral variants. Experimental studies on small subsets of positions have found that landscapes are often rugged, even when high-order nonlinear interactions between sites are comparatively weaker than pairwise additive mutational effects.[33–35] Indeed, even minor epistatic effects can cause enough complexity in the fitness landscape to make evolutionary trajectories unpredictable.[35,36] Recent advances in deep sequencing, gene synthesis and high-throughput screening has permitted more comprehensive exploration of epistasis. For example, fitness landscapes that include large combinatorial complete subsets of sequence space have demonstrated pervasive epistasis across diverse protein backgrounds and folds.[27,37–41] Further, improvements in modelling protein evolution over short and long timescales indicates epistasis, in the form of entrenchment of a fixed mutation or the acquisition of mutations contingent on already fixed mutations, occur gradually over intermediate time-scales, providing insights into how epistasis itself may evolve in a protein.[42]

A recent example in which ruggedness has been described in a sparse and non-local fitness landscape that encompasses the evolutionarily accessible sequence space involves the *lac* operon repressor, LacI, containing context-dependent and independent residues Tyr7, Ty17 and Arg22 at the lac-O binding site (**Figure 2A**).[43] In addition to functional expression and stability, the primary factors influencing the fitness of LacI include its ability to bind both DNA (protein-nucleic acid interaction) and lactose (protein-small molecule interaction) and to transmit the signal of lactose binding to the DNA binding domain *via* conformational change (allostery). Further, the operator that LacI binds is asymmetrical, meaning the dimeric DNA binding domain must recognize and have affinity for two unique DNA half-sites. Characterization of >1,000 extant and reconstructed ancestors of the LacI DBD have shown DNA binding specificity evolves over a highly rugged fitness landscape, characterized by numerous local optima.[43] This ruggedness is due in part to selection pressure of the DBD to bind two unique DNA sequences of the asymmetric operator (**Figure 2 B,C**); the DBD-operator affinity/recognition landscape can be considered as a composite of the affinity/recognition landscapes of the DBD for each asymmetric half-site –yielding a composite landscape more rugged than each component landscape through landscape 'interference' (**Figure 2D**).[43] Similarly, it has been shown that a mutation in the DNA-binding domain may have no significant impact on function by itself, but when combined with a second mutation in the allosteric site, it may lead to a significant loss of function.[44]

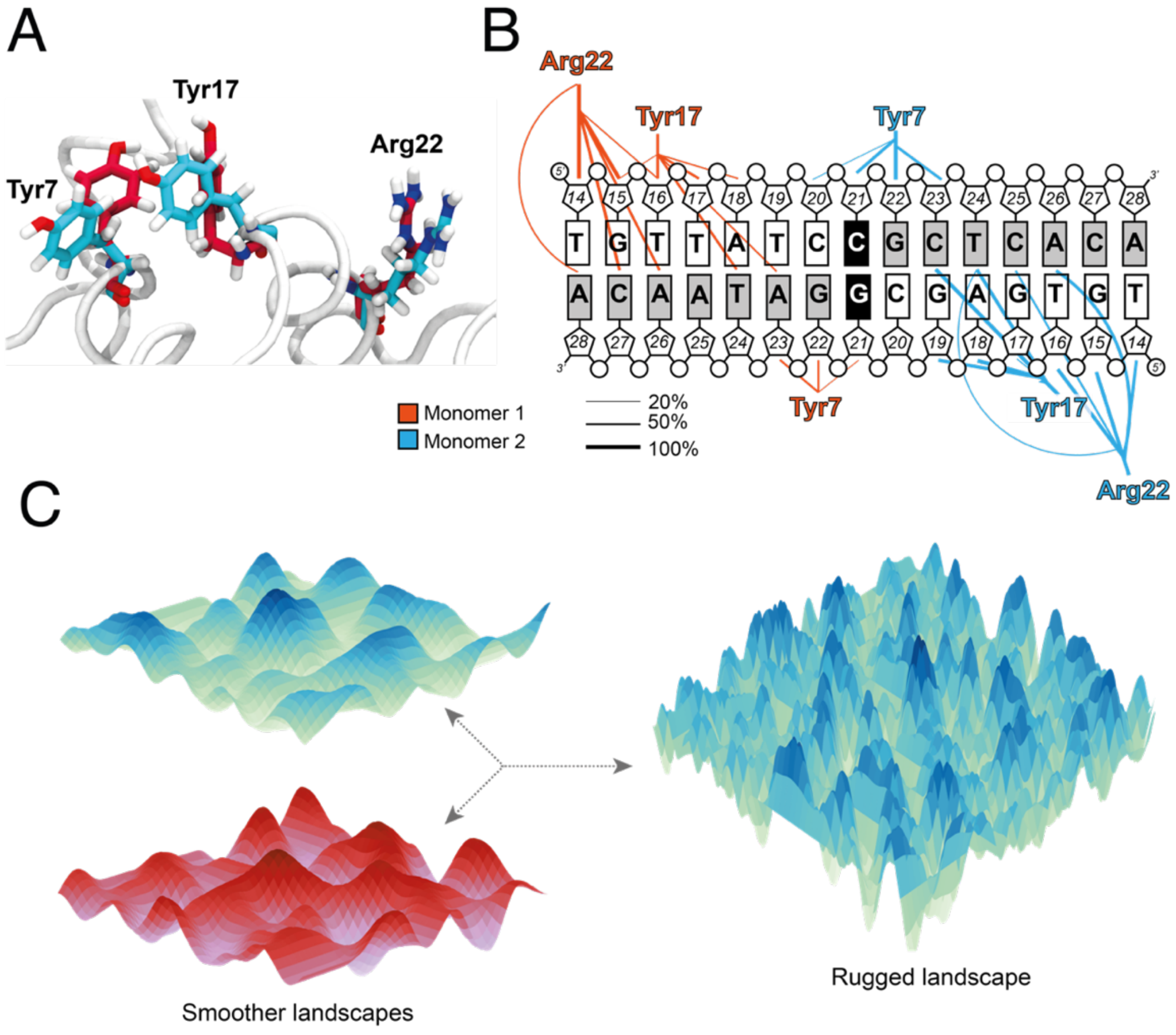

**Figure 2. Composite fitness landscapes can lead to increased ruggedness.** (A) Molecular dynamics model of an ancestral LacI DNA-binding domain (DBD), with each monomer of the LacI homodimer in different conformations (overlaid) when interacting with different DNA half-sites of the DNA operator sequence (not shown).[43] (B) Contact map showing that the DBDs of each monomer in the LacI dimer interact differently with each half-site. Line thickness represents proportion of time interacting. (C) 'Landscape interference', whereby two relatively smooth fitness landscapes (e.g., catalytic activity and thermostability, or binding affinity for two different operator half-sites) can combine to yield a composite landscape that is more rugged than either component landscape.

## Landscape topology as an evolutionary trait

The distinct effects of landscape topology on evolutionary dynamics suggests a hypothesis for the role of fitness landscapes in protein evolution i.e., the ruggedness of the local protein fitness landscape itself could be a trait that is under selection. For example, it has been noted that transcription factors evolve on rugged landscapes, where small changes in the operator sequences can result in drastic changes in fitness.[29,30,43] This may be advantageous for organismal evolution: as transcription factors undergo duplication-divergence and neofunctionalization, fidelity in gene regulation is extremely important; a rugged landscape ensures that there is minimal crosstalk between the evolving effectors of gene expression. In contrast, enzymes, especially those involved in secondary metabolism, are often characteristically promiscuous, with new functions evolving with weak-negative tradeoffs *via* bifunctional intermediates.[45] In this case, catalytic promiscuity can benefit organismal fitness, enabling significant evolution towards a new function prior to duplication and divergence with comparatively little negative selection or deleterious effect of catalytic promiscuity in secondary metabolism. Thus, the characteristics of protein fitness landscapes may themselves be specialized for protein functions.

Evolution may select protein folds for a given role based not only on their immediate fitness, but also their dynamics under the evolutionary process itself. In other words, evolution may act recursively on itself to produce folds that are optimal under mutation and selection based on their general function (e.g. enzymes vs transcription factors). Indeed, the topology of a fitness landscape is dictated by the protein fold and function, which is both heritable and subject to genetic variation. The local fitness landscape topology may therefore be an adaptive trait under selection.[46] That is, given two proteins of different folds but of equivalent fitness, evolution will select the variant with the more favorable fold/fitness landscape, for that function. This second-order/recursive view of protein evolution has important implications for protein engineering, suggesting that protein engineers should select proteins not only based on the initial fitness, but also to optimize the dynamics of evolution on the fitness landscape of the fold. This concept, of different evolvability in different folds, is part of the concept of "fold polarity",[47] and helps explain why a small number of folds have evolved to dominate the sequence space of different functions; for example, the large number of TIM barrel folds among enzymes likely stems from their evolvability for enzymatic reactions because the core scaffold (barrel) has relatively low connectivity to the active site loops, which allows them to adopt a range of conformations and find many solutions to bind substrates and catalyze reactions.

## Traversing rugged fitness landscapes

As noted, the fitness landscapes of real proteins are seldom smooth due to the presence of epistasis[30,31,38] and compounding sub-landscapes.[43] Rugged fitness landscapes present significant challenges for protein evolution, but they also offer opportunities for adaptive innovation.[16] Indeed, evolution manages to navigate the complex interactions of rugged landscapes, circumventing fitness valleys to explore new evolutionary pathways. Likewise, certain features of proteins, like conformational sampling, can help smooth otherwise rugged landscapes. Understanding how proteins overcome the challenges of ruggedness is critical for both natural evolutionary studies and rational protein design, where the goal is often to optimize protein function while avoiding local optima. There are several key mechanisms that facilitate traversal of rugged fitness landscapes, which we now highlight.

*Conformational sampling.* Another important biophysical factor underlying how fitness landscapes are traversed during evolution is conformational sampling, i.e. the ability of proteins to adopt multiple conformational substates.[48,49] Different conformational substates could have different activities and thus contribute differently to the net fitness of the protein: conformation "A" could represent 5% of the populated states, but catalyze a new reaction that the other conformations cannot. Evolution can then act to smoothly "tune" the function of the protein through remote mutations that shift its conformational equilibrium by relatively stabilizing conformation "A". This results in a smoother landscape because there are generally many more pathways by which a pre-existing conformational equilibrium between two conformations can be affected by remote mutations than the number of solutions that are available through complete remodeling of an active site. For example, the laboratory evolution of phosphotriesterase into an into arylesterase was tracked with protein crystallography, which revealed that almost all the mutations were remote from the active site and shifted the conformational equilibrium to favor pre-existing states that were beneficial to the new catalytic activity.[50] The same process was observed in the evolution of a computationally designed Kemp eliminase, wherein gradual changes in the conformational sampling between inactive and active states resulting in a remarkably smooth evolutionary transition.[51] These examples of evolutionary trajectories generated by laboratory evolution demonstrate how conformational flexibility can produce smooth fitness landscapes, and help explain the high frequency of remote mutations observed in evolutionary trajectories.[52]

*Mutational robustness.* The ability of proteins to tolerate mutational changes without losing significant fitness (activity or stability) is referred to as their robustness to mutation. This does not reduce the number of optima, but reduces the number and severity of fitness valleys, making the fitness landscape easier to traverse. For example, a functionally important structural trait could be achieved through multiple independent structural innovations e.g., a mobile loop could be stabilized by a hydrogen bond to one region of the protein, and/or a hydrophobic pocket that attracts and stabilizes a nearby hydrophobic sidechain.[50] Deleterious mutational effects can also be buffered by other factors. At the protein level for example, a thermostability "buffer" can offset stability losses of a mutation that would have led to unfolding.[53–55] Likewise, environmental factors, such as the presence of chaperones, can alter the shape of the landscape and make it smoother by allowing mutations that would normally destabilize the protein to be incorporated, creating more pathways to fitness peaks.[56]

*Compensatory and contingent mutations.* Compensatory mutations are mutations that offset the negative effects of prior mutations, allowing proteins to escape local optima and explore new regions of sequence space.[57] For example, in the case of LacI, a mutation that reduces its ability to bind DNA can be compensated for by a second mutation in a distant region of the protein, restoring its DNA-binding ability through changes in conformation or flexibility.[58] A related concept is contingency, where certain mutations occurring beforehand permit the traversal of otherwise non-viable trajectory by mitigating otherwise deleterious effects.[59] The distinction between compensation/contingency and global buffering of protein properties that leads to mutational robustness, is that compensation and contingency are subject to the trajectory being taken. That is, in the absence of other buffers, an otherwise deleterious mutation can be incorporated in an evolutionary trajectory if compensatory mutations are accessible, or contingent mutations are present.

*Neutral networks.* Adaptive walks, the series of mutations that a protein undergoes as it climbs toward higher fitness, are often constrained by ruggedness.[60] To overcome these barriers, proteins must sometimes accumulate neutral mutations—mutations that do not immediately affect fitness significantly but create new opportunities for adaptation by altering the underlying structural or functional framework of the protein.[10,20] For example, work on the evolution of beta-lactamase, an enzyme that provides resistance to antibiotics, showed that many of the most direct paths to high fitness were blocked by mutations that reduced fitness,[61] creating a fitness valley in the landscape. In these cases, the enzyme had to rely on compensatory mutations to maintain functionality while traversing the fitness valleys,[55] highlighting the complexity of adaptive walks in rugged landscapes. Indeed, neutral drift can enable proteins to escape local optima and, over time, neutral mutations can accumulate and lead the protein to new regions of the fitness landscape creating new opportunities for adaptation where adaptive walks to different local fitness maxima become accessible.[62–64]

## A theoretical framework for ruggedness

Epistasis and ruggedness have been described mathematically using tools from discrete mathematics and graph theory, permitting a more comprehensive understanding of fitness landscapes, their properties, and the characteristics of evolutionary processes that unfold on them.[65–67] Representing fitness landscapes as network graphs enables the study of landscape topology within established mathematical theory and frameworks. In a fitness landscape network graph, sequences are encoded as nodes and edges connect nodes based on sequence similarity. The simplest scheme is the Hamming graph, where sequences (nodes) are connected to their neighbors based on the number of sequence differences (mutational distance) e.g., nodes could be connected by edges to other nodes by a Hamming (mutational) distance of 1, or more **(Figure 3A)**.

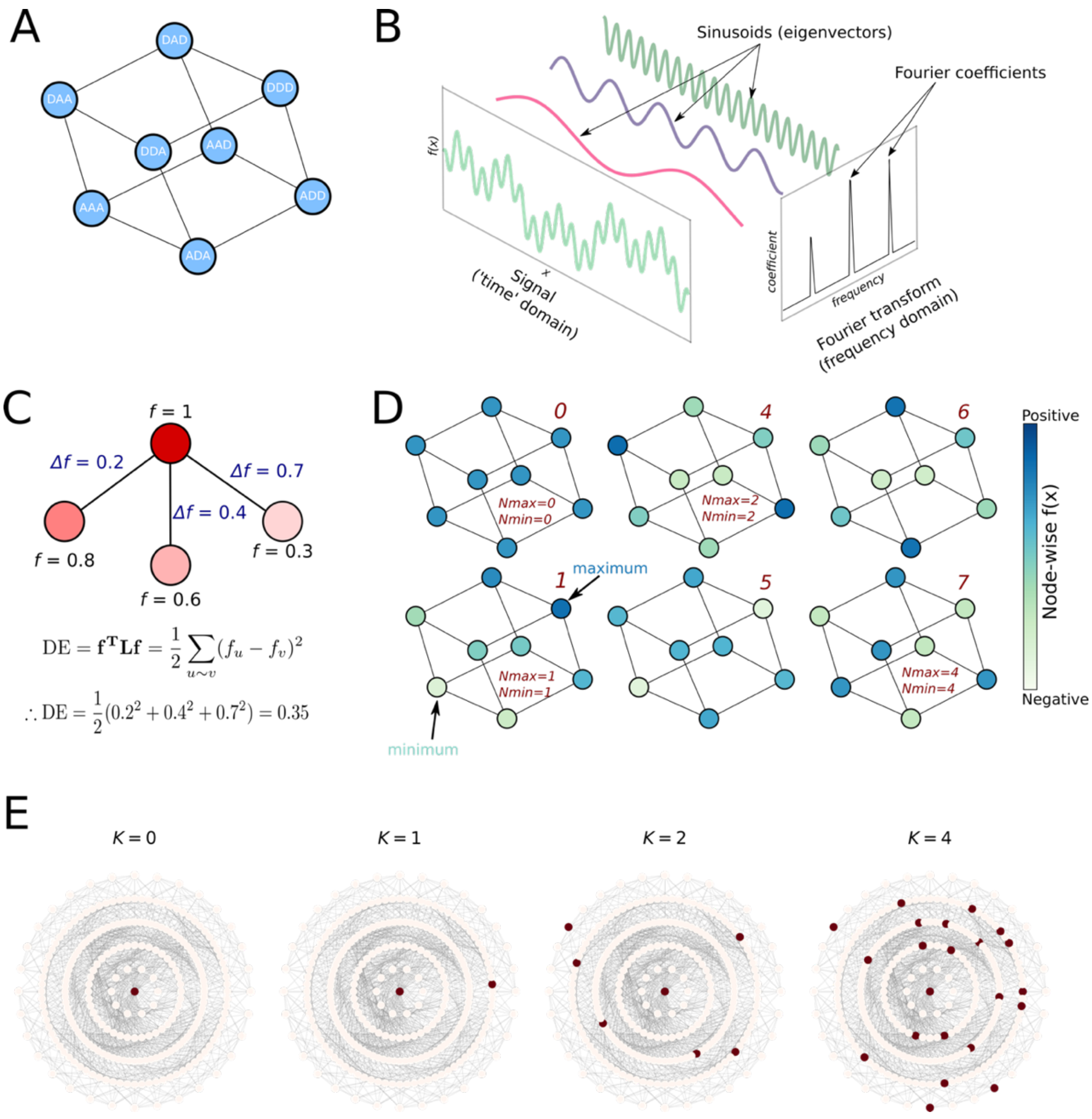

$$\mathrm{DE} = \mathbf{f}^{\mathbf{T}}\mathbf{L}\mathbf{f} = \frac{1}{2}\sum_{u\sim v}(f_u - f_v)^2$$

$$\therefore \mathrm{DE} = \frac{1}{2}(0.2^2 + 0.4^2 + 0.7^2) = 0.35$$

**Figure 3. Representations of protein fitness using discrete mathematics and graph theory**. (A) The graph representation of a sequence space. Here, nodes represent a sequence, and edges connect sequences to their single-mutation neighbors (i.e., Hamming graph). (B) The continuous Fourier transform is a technique for expressing a signal *f(x)* as a linear combination of orthogonal sinusoids (eigenvectors) of increasing frequency. The Fourier transform of the signal *f(x)* represents the coefficients of the linear combination. Fourier transforms of complex signals have more contribution from higher-frequency sinusoids than less complex signals. (C) The Dirichlet energy of signal *f(x)* over a graph is a measure of the variability (ruggedness) of *f(x)*. Nodes denoting sequences *x* are shown with corresponding fitness values *f(x)*, and *Δf* denotes the edge-wise gradient in fitness between connected nodes. The Dirichlet energy is then calculated as a squared-sum of these edge-wise gradients. **f** denotes the vector of fitness values over the graph, **f^T** its transpose, and **L** denotes the graph Laplacian. (D) Eigenvectors of increasing 'frequency' the graph from panel A are shown; color bar shows the node-wise value of *f(x),* where *x* is the graph domain. (E) Radial representation of graphs of NK landscapes of increasing complexity, with local maxima shown in red. As the value of the *a priori* epistasis parameter *K* increases, the number of local maxima also increases.

Fourier analysis serves as a valuable method for breaking down fitness landscapes into their fundamental epistatic components **(Figure 3B)**.[68,69] Applying this technique to the fitness landscape yields Fourier coefficients that can be interpreted as the contribution of interactions of different epistatic orders to the overall fitness.[70] This decomposition can provide a clearer understanding of the fitness relationships that emerge from various combinations of mutations since each coefficient in the Fourier transform corresponds to a specific epistatic interaction - helping us map the ruggedness of the landscape and the complex fitness effects that guide evolutionary trajectories. In this way, Fourier analysis uncovers higher-order interactions essential for understanding how proteins evolve and adapt. By representing the fitness landscapes as Hamming graphs, where the nodes represent different sequences and edges connect sequences to their neighbors, we can use the Graph Fourier Transform (GFT) to expedite the Fourier analysis.[71] The GFT expresses fitness as a linear combination of the graph's Fourier basis, obtained from the eigenvectors of the graph's Laplacian matrix, enabling efficient analysis of signals on these graphs. This extends the use of GFTs in graph signal processing[72] to the analysis of protein fitness landscapes **(Figure 3D)**.

Another approach to quantify the extend of ruggedness is to use the Dirichlet energy of a network graph, which offers a more interpretable measurement of local and global ruggedness when the underlying data structure is sparse **(Figure 3C)**.[43,73,74] The Dirichlet energy describes how variable a function is: for a given fitness function, high Dirichlet energy indicates that fitness changes rapidly between neighboring sequences, signaling a rugged landscape. This method has been used for the quantification of the global ruggedness, as well as local ruggedness of specific regions of the fitness landscape, providing insights into how epistasis constrains protein evolution.[43,74–76]

By using well-established fitness landscape models, researchers can control *a priori* ruggedness and thereby test how well different measures of ruggedness perform in recapitulating the underlying complexity of the fitness landscape. One of the most widely used fitness landscape models is the NK model.[77] In the NK model, N denotes the number of sites in a protein, while K represents the average number of interacting sites. When K=0, each site contributes independently and additively to fitness, and the landscape is smooth. However, as K increases, the degree of epistasis increases, meaning each site's contribution to overall fitness is dependent on the presence/absence of particular mutations at other sites, and the landscape becomes more rugged, with numerous local optima **(Figure 3E)**. This model has provided valuable insights into how proteins navigate rugged fitness landscapes through both natural evolution and experimental directed evolution. It also offers valuable insights into the effects of ruggedness on ML performance, indicating that performance decreases as ruggedness increases.

Despite its utility, the NK model has limitations in terms of capturing the nuances of real-world protein evolution.[71] This has led to variations of the model. The generalized NK model can be used to identify higher-order epistatic interactions in the fitness functions,[71] the NKp family accounts for neutral networks by assigning no fitness contributions to some mutations,[78] and structurally informed NK models can model realistic local-epistatic interactions.[71] Despite this, many phenomena that occur in proteins are challenging for NK-like models. Distal interactions due to interaction networks,[48,79] conformational dynamics and allostery,[80,81] location-dependence of mutation impact,[82,37] and the presence of functional hotpots[83] are difficult to model despite the overall fitness landscape arising from an interplay of these non-random factors among others.

## Machine Learning and Fitness Landscape Prediction

Machine learning (ML) has become a valuable tool in the study and optimization of protein fitness landscapes, particularly in handling the complexity and ruggedness that characterize many of these landscapes.[12,84–86] By integrating data-driven models with computational techniques, ML enables researchers to predict fitness outcomes from protein sequence variations more efficiently, guiding both experimental evolution and rational protein design. One of the biggest challenges lies in navigating rugged fitness landscapes, where epistasis and the presence of local optima make traditional evolutionary methods slow and unpredictable. Recent ML advancements, including neural networks, graph-based models, and generative techniques, have significantly enhanced the capacity to map and explore these landscapes.

The use of neural networks, particularly deep learning models, to model the relationship between protein sequence and function has proven to be powerful.[87,88] Multi-layer perceptrons (MLPs) are among the simplest neural network approaches, yet have been shown to be versatile and powerful tools to predict fitness of sequences from relatively sparse datasets.[89] Likewise, convolutional neural networks (CNNs), have been shown to predict fitness across sequence spaces,[87] albeit with varying degrees of success depending on the architecture used. Indeed, simple ensembles of convolutional neural networks were found to excel in local extrapolation, while more complex models could capture nonlinear interactions essential for rugged fitness landscapes. However, CNNs tend to struggle to extrapolate when sequences deviate significantly from the training data.[87] Graph-based methods have also proven effective, particularly in smoothing rugged landscapes. A recent report of Gibbs Sampling with Graph-based Smoothing (GGS) was shown to identify local optima by applying graph-based smoothing techniques to fitness data.[90] This method enables better fitness predictions by mitigating the noisy, rugged nature of these landscapes. Such smoothing techniques are particularly valuable when dealing with data-scarce environments, where limited experimental data can lead to overfitting or poor generalization in traditional ML models.

Protein language models, which learn expressive protein representations in their latent spaces through masked language tasks,[91–93] have perhaps seen the most significant advancements in recent years and offer the most flexible functionalities in learning fitness landscape topologies. For example, fine-tuning pretrained protein language models to map sequence likelihoods to empirical fitness scores has demonstrated significant progress towards learning landscape structure from sparse experimental data.[94,95] Combining large protein language models with reinforcement or genetic search algorithms can also offer effective avenues to both learn and traverse the structure of the fitness landscape.[12,96,97]

The value of smoothing was also exemplified in the Learned Ancestral Sequence Embedding (LASE) technique, which was shown to smooth rugged fitness landscapes by leveraging evolutionary priors (the assumed probability distribution).[76] The LASE method enables ML models to learn smoothed representations of the fitness landscape, thus improving their ability to predict high-fitness variants even when the actual landscape is rugged with multiple peaks and valleys. The smoothing provided by LASE reduces the influence of epistasis and local optima, facilitating more efficient exploration of sequence space. Similar approaches have imposed ruggedness- and epistasis-specific regularization to dictate the deep representation model learns a mapping to a smooth fitness landscape structure latent representation.[75,98]

Finally, generative models, such as variational autoencoders (VAEs), have also been applied to predict and generate protein sequences with high fitness.[99] These models are particularly valuable in experimental setups with limited data, as they can efficiently explore large sequence spaces and suggest optimal candidates for further mutagenesis or screening. VAEs, for example, compress the sequence space into a lower-dimensional representation, making it easier to visualize and explore fitness landscapes. Indeed, such VAEs have been shown to be effective in the prediction of disease variants by leveraging evolutionary data.[100]

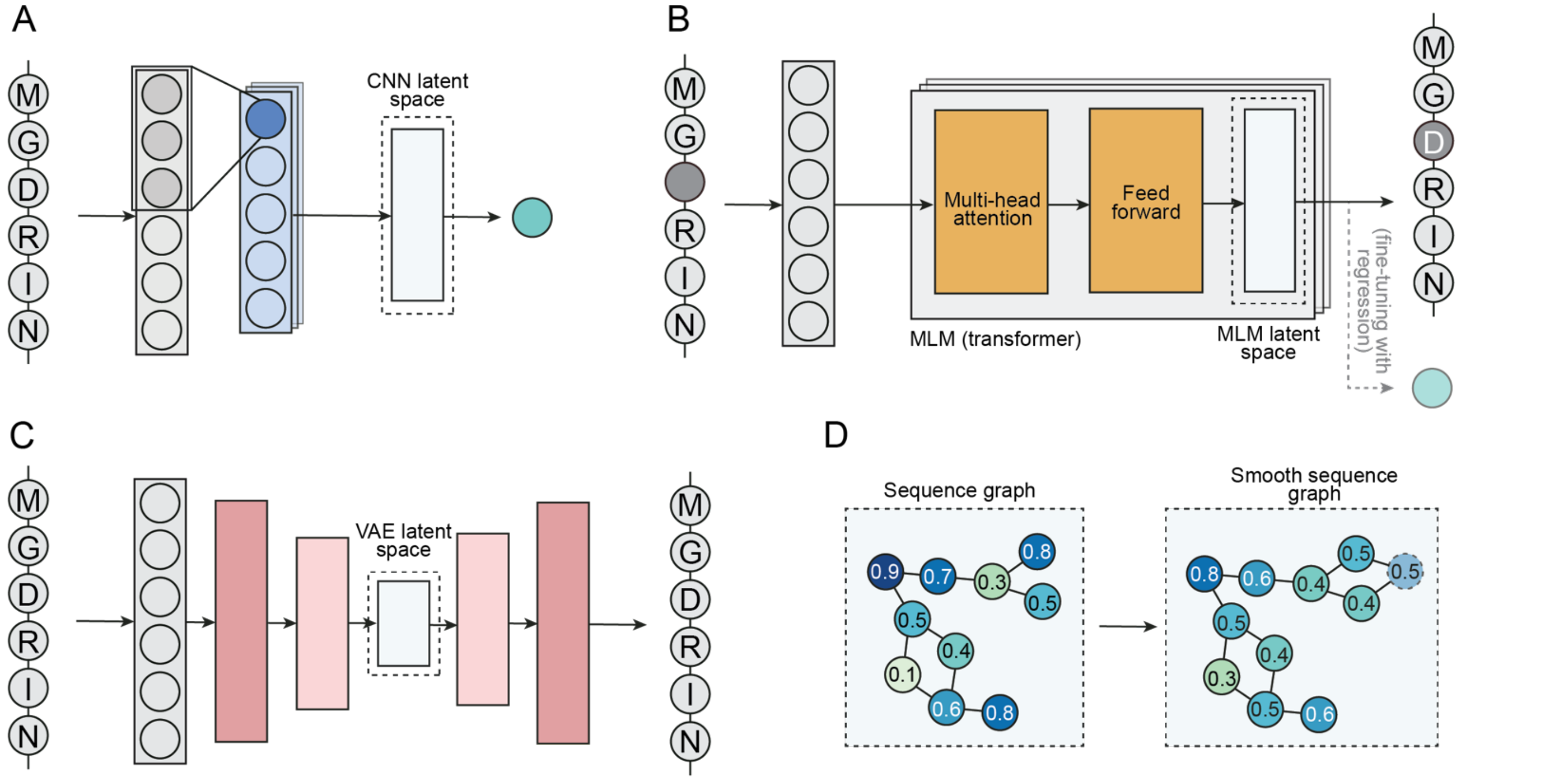

**Figure 4. Machine learning methods for producing protein representations and predictions.** (A) Machine learning with a convolutional neural network (CNN) architecture to form a latent/representation space prior to prediction of protein fitness. (B) Masked language modelling (MLM) with a transformer architecture to learn latent/representation space prior to prediction of masked amino acids. The architecture can be fine-tuned further to predict protein fitness. (C) Sequence compression into a low-dimension latent/representation space by implementing variational auto encoder (VAE) that can still be reconstructed. (D) Illustration of Graph-based Smoothing to smooth a sequence-fitness graph and generate augmented datapoints (dotted-line circle).

## Conclusion and future directions

Fitness landscapes provide a powerful conceptual framework for understanding how proteins evolve and for guiding protein engineering efforts. Whether dealing with smooth or rugged landscapes, the topography of the landscape determines the ease with which proteins can evolve or be engineered. In smooth landscapes, evolutionary pathways are relatively straightforward, with mutations leading to predictable improvements in fitness. In contrast, rugged fitness landscapes present significant challenges, as multiple local optima and fitness valleys create barriers to optimization. Proteins in rugged landscapes must navigate a more complex evolutionary process, often requiring compensatory mutations, adaptive walks, and neutral drift to escape local optima and reach higher fitness peaks. Epistasis plays a critical role in shaping rugged landscapes, leading to complex, non-linear interactions between mutations that make it difficult to predict evolutionary trajectories.

Despite these challenges, data generated by directed evolution and ancestral sequence reconstruction have improved our ability to explore and exploit fitness landscapes for protein optimization. Directed evolution mimics natural selection in the laboratory, allowing researchers to generate large libraries of protein variants and identify those with enhanced functions.[11] In smooth landscapes, directed evolution can proceed predictably, but in rugged landscapes, additional strategies such as recombination and compensatory mutations are often required to overcome local fitness peaks and experiments can frequently plateau or “stall”. Ancestral sequence reconstruction has allowed for much wider, albeit sparse, sampling of the sequence space and fitness landscape.[101,102] Together these approaches have generated many of valuable datasets that we can now learn from using advanced computational approaches.

The integration of machine learning has further revolutionized the study and prediction of protein fitness, providing predictive models that can generalize (with various levels of success) from small datasets to infer the fitness of untested variants. These models are particularly valuable and complement directed evolution in the case of rugged landscapes, where the effects of mutations are non-linear and difficult to predict. By combining machine learning predictions with high-throughput screening and deep mutational scanning, researchers can explore large regions of sequence space more efficiently, increasing the likelihood of discovering new fitness peaks. Indeed, neural networks, graph-based models like GGS, smoothing techniques such as LASE, and generative methods like VAEs are revolutionizing the study of protein fitness landscapes. By smoothing landscapes and mitigating the effects of noise, ML models are now able to better guide the design and evolution of proteins with improved functionality.

The future of protein engineering will likely see even more sophisticated ML techniques become integrated with experimental methods like directed evolution. As our understanding of fitness landscapes deepens, so too will our ability to design proteins with novel or enhanced functions. By combining traditional experimental techniques with cutting-edge computational tools, we can better navigate the complex topographies of fitness landscapes, optimizing proteins for a wide range of applications in biotechnology, medicine, and industrial processes. The future of protein design arguably lies in the integration of empirical data, computational models, and machine learning, enabling us to harness the full potential of protein evolution.